\documentclass{rnaastex}
\usepackage{txfonts}

\begin{document}
\title{\Large Kinetics and clustering of dust particles in supersonic turbulence with self-gravity}
%% Note that the corresponding author command and emails has to come
%% before everything else. Also place all the emails in the \email
%% command instead of using multiple \email calls.
\correspondingauthor{Lars Mattsson}
\email{lars.mattsson@su.se}
\author{Robert Hedvall}
\affiliation{Department of Physics, Stockholm University, Stockholm, Sweden}
\author{Lars Mattsson}
\affiliation{Nordita, KTH Royal Institute of Technology \& Stockholm University, Stockholm, Sweden}

%% Note that RNAAS manuscripts DO NOT have abstracts.
\vspace*{-0.9 cm}

%% Start the main body of the article. If no sections in the 
%% research note leave the \section call blank to make the title.

\section{Hydrodynamic simulations with particles} 
We present a simulation of isothermal supersonic (rms Mach number $\mathcal{M}_{\rm rms}\sim 3$) turbulent gas with inertial particles (dust) and self-gravity in statistical steady-state (S3G), which we compare with a corresponding simulation without self-gravity (S3), similar to those in \citet{Mattsson19a}. The computational domains are 3D boxes ($512^3$) with periodic boundaries and sides of equal length $L = 2\pi$, where we solve the compressible Navier-Stokes equations with stochastic forcing. We use the  {\sc Pencil Code}, a high-order finite difference code capable of simulating compressible flows with inertial particles \citep{Brandenburg02}. 
%Applied to, e.g., a molecular cloud (MC) core, the physical size of the box is about $L=0.1-0.5$ pc which yields a resolution of $\sim 40-200$ AU. The core of an MC would also have a mean number density of gas particles in the range $10^3-10^5$ cm$^{-3}$. %The time scale is set to be the time it takes for sound to cross the length of the simulation box i.e., the sound-crossing time $\tau_{\rm sc}=L/c_{\rm s}$. Since $c_{\rm s}$ is set to unity $\tau_{\rm sc}=L$.

The gas flow is characterized by the density $\rho$ and the flow velocity $\mathbf{u}$. Dust-particle velocities $\mathbf{v}_i$ are followed using a Lagrangian equation of motion (EOM) for each particle $i$ ($10^7$ discrete inertial particles in 10 size bins with $10^6$ particles). Grain-sizes are charachterized by $\alpha = ({\rho_{\rm gr}}/{\langle \rho \rangle})\,({a}/{L})$,where $a$ is the grain radius and $\rho_{\rm gr}$ is the bulk material density \citep{Hopkins16}. Without self-gravity, the simulation can thus be scaled arbitrarily. S3G, however, is not scale-free. It is in steady state, but close to gravitationally unstable, since we adopt a {\it Jeans wavelength}, $\lambda_{\rm J}=2\pi = L$, which provides the strongest influence of gravity on the dynamics of gas and dust without causing irreversible gravitational collapses. 
%The S3 simulation share the same basic setup, except we have no self-gravity.
%, so there is no perturbation wavelength or box size sensitivity to consider. 

%\section{Results and implications for further study}
\vspace*{0.9 cm}
\section{Results and directions for further study}
\subsection{Mean velocities and clustering}
Clustering of particles can be quantified by their first nearest neighbor distance (1-NND) \citep{Monchaux12}. We compare the measured average 1-NND $\langle r_{\rm obs} \rangle$ to that expected from a random isotropic distribution $\langle r_{\rm exp} \rangle$. This average nearest neighbor ratio, $R_{\rm ANN}={\langle r_{\rm obs} \rangle}/{\langle r_{\rm exp} \rangle}$, is $R_{\rm ANN} \approx 1$ for unclustered grains. If $R_{\rm ANN} < 1$ the particles are clustered, since the average 1-NND is lower. We must distinguish between {\it compaction} and {\it fractal} clustering, where the latter is due to rotation of the flow rather than compression. Fractal clustering is usually measured by the correlation dimension $d_2 = \lim_{\delta r\to 0} \left\{\ln[\langle \mathcal{N}(\delta r)\rangle]/\ln \delta r\right\}$, where $\mathcal{N}$ is the expected number of particles inside a ball of radius $\delta r$ surrounding a test particle at the centre of the ball.

Fig. \ref{fig} (lower left) shows that root-mean-square (rms) grain velocities depend on particle size, since lighter particles couple better to the gas due to lower inertia than heavier particles. $R_{\rm ANN}$ and $d_2$ (Fig. \ref{fig}, left panels) are also determined by particle size, except for $\alpha = 8.0,\ 16.0$. Particles coupled to a turbulent gas will experience a higher degree of fractal clustering (low $d_2$) since they get temporarily trapped in vortices and may therefore accumulate in convergence zones in between vortices as they are being expelled \citep{Mattsson19a}. 

\subsection{The influence of gravity}
\begin{figure}[h!]
\begin{center}
\resizebox{\hsize}{!}{
\includegraphics{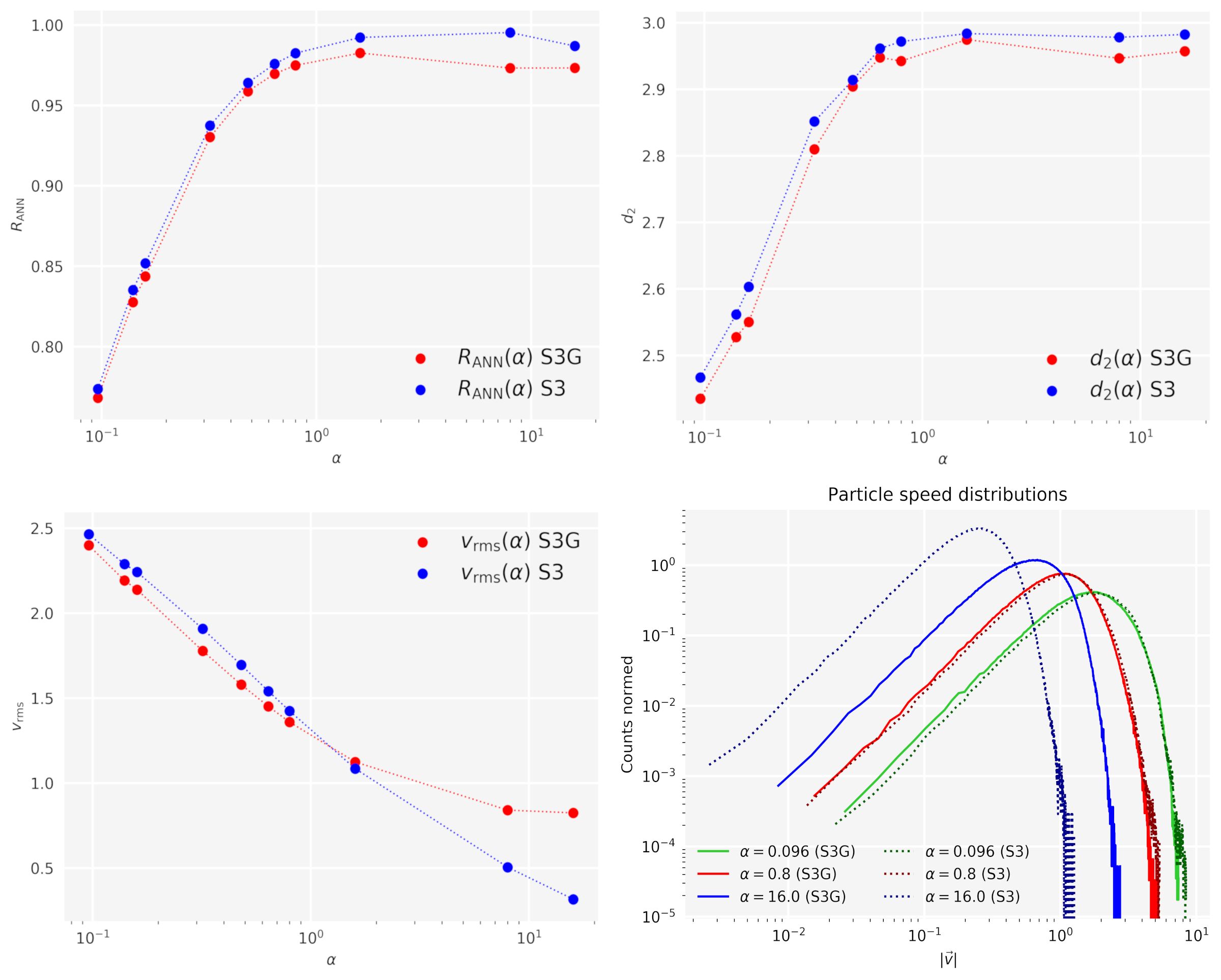}}
%\vspace*{-0.7 cm}
\caption{\label{fig} Comparison of simulations. {\it Top}: $R_{\rm ANN}$ and $d_2$ as functions of $\alpha$. {\it Bottom}: $v_{\rm rms}$ in units of $c_{\rm s}$ as function of $\alpha$ and speed distributions for different $\alpha$. }
\end{center}
\end{figure}

Comparing S3G and S3, we find that self-gravity does not cause any significant increase in clustering, as determined by $R_{\rm ANN}$ and the correlation dimension ($d_2$), regardless of particle size (Fig. \ref{fig}, upper panels). However, there is a brief initial phase of strong clustering for $\alpha = 8.0,\ 16.0$ in both simulations, but much more prominent in S3G where particles are also gravitationally accelerated. With more inertia, and thus less affected by drag, heavier particles free-fall and cluster faster than lighter particles coupling to the turbulent gas. 

When a steady state is reached, $v_{\rm rms}$ is a function of $\alpha$, which can be derived from the EOM,
\begin{equation}
\label{eom}
\frac{d\mathbf{v}_i}{dt}=\frac{\mathbf{u}-\mathbf{v}_i}{\tau_{{\rm s},\,i}}-\nabla {\Phi},
\end{equation}
where $\Phi$ is the gravitational potential and the stopping time $\tau_{{\rm s},\,i}$, defining the coupling efficiency between gas and dust particles, is given by \citep{Draine79},
\begin{equation}
\label{stoptime}
\tau_{{\rm s},\,i} = \sqrt{\frac{\pi}{8}}\frac{\rho_{\rm gr}}{\rho}\frac{a}{c_{\rm s}}\left( 1+ \frac{9\pi}{64}\frac{|\mathbf{v}_i-\mathbf{u}|^2}{c^2_{\rm s}} \right) ^{-1/2}.
\end{equation}
Note that $\tau_{{\rm s},\,i} \propto \alpha$ explains why there is no significant change in velocities for the small particles (kinetic drag dominates). Fig. \ref{fig} (lower left) shows that only heavy particles ($\alpha =8.0,\ 16.0$) show elevated $v_{\rm rms}$ in the presence of self-gravity and the speed distributions (lower right) are significantly affected for $\alpha =8.0,\ 16.0$ (distributions are shifted to higher values in S3G compared to S3). For $\alpha =8.0$ the rms-velocity increases from $v_{\rm rms}/c_{\rm s} = 0.52$ to 0.84 (62\%) and for $\alpha =16.0$, $v_{\rm rms}/c_{\rm s} = 0.33$  rises to 0.83 152\%). 
%These increments are probably the combined result of kinetic drag and gravity. 

%\vspace*{-0.1 cm}
The stopping time $\tau_{{\rm s},\,i} \propto \alpha$ differs by a factor of $\sim 160$ between the smallest and the largest particles. Acceleration due to gravity becomes the dominant force for sufficiently large particles, which explains the discrepancies seen in Fig. \ref{fig}. To describe the difference in velocities between S3G and S3, we note that $v_{\rm rms}$ satisfies a realtion $v^2_{\rm rms}/u^2_{\rm rms} =1  - \Psi(\alpha)$, with $\Psi = \langle (\mathbf{u} + \mathbf{v}_i)\cdot (\mathbf{u} - \mathbf{v}_i) \rangle / \langle \mathbf{u}\cdot  \mathbf{u} \rangle$. Assuming $\tau_{{\rm s},\,i}$ is a function of $\alpha$ only, (\ref{eom}) and statistical steady-state yields,
%\begin{equation}
%\Psi(a) = {1\over u_{\rm rms}^2}\left\langle (\mathbf{u} + \mathbf{v}_i)\cdot \left( {{\rm d} \mathbf{v}_i\over {\rm d} t} + \nabla \Phi \right) \tau_{{\rm s},\,i}\right\rangle.
%\end{equation}
\begin{equation}
\Psi  = {\tau_{{\rm s},\,i}^2\over u_{\rm rms}^2}\left(\left\langle {{\rm d} \mathbf{v}_i\over {\rm d} t}\cdot {{\rm d} \mathbf{v}_i\over {\rm d} t} \right\rangle  - \langle \nabla \Phi \cdot \nabla \Phi \rangle \right),
\end{equation}
where terms within brackets are squares of the rms values of total and gravitational acceleration ($a_{\rm rms}$ and $g_{\rm rms}$, respectively). First, we note that in the tracer-particle limit (very small grains; $\mathrm{v}_i\to \mathrm{u}$), $\Psi = 0$. Second, for larger particles, $v_{\rm rms}$ as a function of $\alpha$ must reach a minimum and then rise again because $g_{\rm rms} = {\rm constant}$ and $a_{\rm rms}$ must be a bounded function of $\alpha$. Thus, $\Psi$ must also be a bounded function. For large grains, $\tau_{{\rm s},\,i}$, as in (\ref{stoptime}), is actually limited by $|\mathbf{u}-\mathbf{v}|/c_{\rm s}$ and $\Psi$ will approach an arbitrary finite value as $\tau_{{\rm s},\,i}\,a_{\rm rms}/\tau_{{\rm s},\,i}\,g_{\rm rms}$ approaches its upper limit. Obviously, $\mathcal{M}_{\rm rms}$ plays a role here and for $\mathcal{M}_{\rm rms} \sim 1$ simple scalings suggest that $v_{\rm rms}$ reaches its minimum for $\alpha \sim 1$ and approach a finite value for really large grains, possibly exceeding $u_{\rm rms}$. More and extended simulations will likely confirm or disprove this idea.

%\acknowledgments

%\vspace*{-0.9 cm}
\bibliographystyle{apj}
\bibliography{refs_dust}

\end{document}